\documentclass[preprint,12pt]{elsarticle}

\usepackage{graphicx}
\usepackage{subfigure}
\usepackage{amssymb}
\usepackage{amsmath}
\usepackage{cases}
\usepackage{color}

 \usepackage{lineno}

\journal{a Journal}

\begin{document}

\begin{frontmatter}



\title{Initial Ranging for Prioritized Network Entry in \\IEEE 802.16 Network}


\author{Subodh Pudasaini}
\author{Seokjoo Shin}
\address{Department of Computer Engineering, Chosun University, Gwangju, Korea}


\begin{abstract}
Prioritized network entry is desirable for establishing preferential network connectivity for the higher priority users when different priority users exist over a given network. In line with such desirability, we propose a simple but efficient priority differentiated initial ranging mechanism considering an Orthogonal Frequency Division Multiple Access (OFDMA) based IEEE 802.16 network. In the proposed mechanism, we introduce an approach that integrates an explicit CDMA-ranging code reservation scheme with a Ranging Slot Selection Window (RSSW) differentiation scheme. Simulation results are provided to characterize the performance of the proposed mechanism.

\end{abstract}

\begin{keyword}

Initial ranging \sep Prioritized network entry \sep IEEE 802.16


\end{keyword}

\end{frontmatter}

\linenumbers

\section{Introduction}
Network entry in wireless mobile communication systems is a set of procedures that a Mobile Station (MS) should follow in order to enter the operator's network and obtain network services. These procedures configure network connectivity parameters, achieve uplink synchronization, and enable power control for MSs \cite{Ahmadi} \cite{Taha} \cite{Andrews}.

In Orthogonal Frequency Division Multiple Access (OFDMA) based IEEE 802.16 network \cite{standard}, the network entry process begins with a MS sending a predetermined ranging code (at a calculated transmit power level) in a randomly selected ranging slot once the down link synchronization with Base Station (BS) is achieved. If it does not receive any ranging response from the BS within certain time duration, then the MS considers its previous ranging attempt to be unsuccessful and enters the contention-resolution stage. The MS then sends a new ranging code at the next ranging opportunity with a one-step-higher power level. This process continues until the MS receives a ranging response message indicating the completion of the process. The MS may now start its uplink transmission after negotiating the basic capabilities such as physical and bandwidth allocation related parameters.

IEEE 802.16 has standardized a Quality of Service (QoS) differentiated resource allocation framework \cite{Cicconetti} \cite{Zhang}. Such a framework allows implementation of class based resource scheduling algorithms wherein the channel resources can be distributed to MSs (only to those which have already made successful network entry) relative to the priority of their traffic. However, it does not specify any mechanism for providing prioritized network entry. The concept of prioritized network entry is becoming more relevant in context of Emergency Telecommunication Services (ETS) over the emerging telecommunication infrastructures because such services should not only be allowed to receive preferential use of operational network resources but should also be provided with higher assurance of maintaining network connectivity over the non-emergency regular services \cite{Folts}\cite{Boone}. To this end, we present a priority differentiated network entry mechanism in context of IEEE 802.16 based telecommunication networks.

The proposed mechanism incorporates two different approaches for access prioritization of high priority MSs. The first approach, which is inspired from the QoS differentiation framework in wireless local area networks \cite{Zhu}, differentiates the size of starting Ranging Slot Selection Window (RSSW) for high priority and low priority MSs while the second approach explicitly reserves a certain fraction of available number of ranging codes to high priority MSs. The proposed scheme has also been under discussion in IEEE 802.16 working group \cite{weblink}.

The rest of the paper is organized as follows: Section $2$ provides preliminaries on ranging process and initial ranging while the following Section $3$ describes the proposed priority initial ranging mechanism. Section $4$ provides simulation results. Finally, Section $5$ gives conclusion of the work described in this paper.

\section{Preliminaries}
In this Section, based on the presentation in \cite{Ahmadi}\cite{Taha}, we present a quick overview of network entry and ranging mechanism in IEEE 802.16-2012 considering the OFDMA PHY. Note that initial ranging mechanism for Single Carrier FDMA (SC-FDMA) and OFDM PHYs is slightly different.

\subsection{Network Entry}
Fig. 1 depicts the network entry procedure in IEEE 802.16-2012 that a MS should follow in order to enter the operator's network and obtain network services. Upon initialization or powering up, an MS first scans the downlink (DL) frequencies to find whether it is within the coverage of a BS. Once a DL preamble is detected, the MS synchronizes itself with respect to the DL transmission of the BS.

\begin{figure}[h]
\centering
\includegraphics[width=80mm, height=100mm]{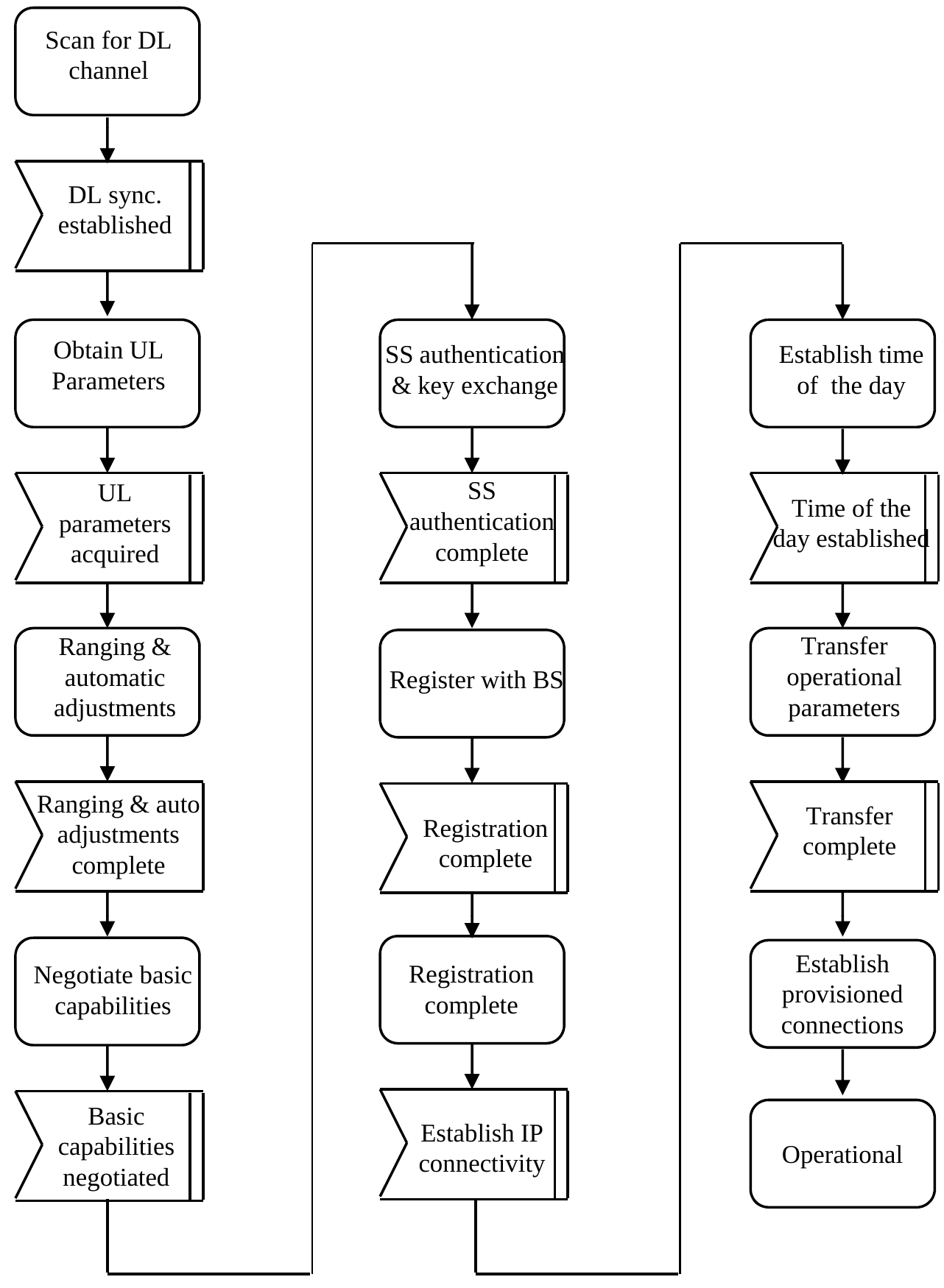}
\centering
\caption{Flowchart of network entry and initialization in IEEE 802.16-2012 \cite{standard}}
\end{figure}

\noindent As the DL synchronization is achieved, the MS then listens to the various control messages (such as Frame Control Header (FCH), DL channel Descriptor (DCD), Uplink Channel Descriptor (UCD), DL-Allocation map (DL-MAP) and UL-allocation map (UL-MAP))  that follow the DL preamble to obtain the various PHY as well as MAC related  parameters for DL and UL transmissions. From the UL-MAP the MS gathers necessary information about the ranging opportunities and perform ranging operation. Once ranging is successfully completed, the MS negotiates its basic capabilities regarding various PHY and bandwidth allocation related parameters. If authorization is enabled in the network, the MS performs authorization and key exchange and then attempts to register itself with the BS. It would then proceed to establish IP connectivity. The subsequent optional stages of establishing time of the day and transferring operational parameters are then pursued if needed. The MS could now establish provisioned connections and is considered operational. Its operational status thereafter is maintained through periodic ranging.

\subsection{Initial Ranging}
Ranging is an uplink physical layer procedure that maintains the quality and reliability of the radio-link communication between BS and MS. The ranging process involves a transmission of a deterministic sequence known as ranging code over a number of OFDM symbols in the ranging channel. The BS processes the received signal on the ranging channel to estimate the several radio-link parameters including channel impulse response, Signal to Interference Noise Ratio (SINR), and time of arrival, that are used for timing/frequency/power adjustments of the uplink transmissions from the MS. MSs cannot perform any uplink transmissions until they successfully complete the ranging procedures. The ranging process is classified into two categories: (a) initial ranging for MSs that are not uplink-synchronized, and (b) periodic ranging for MSs that are already uplink-synchronized.

\begin{figure}[h]
\centering
\includegraphics[width=50mm, height=70mm]{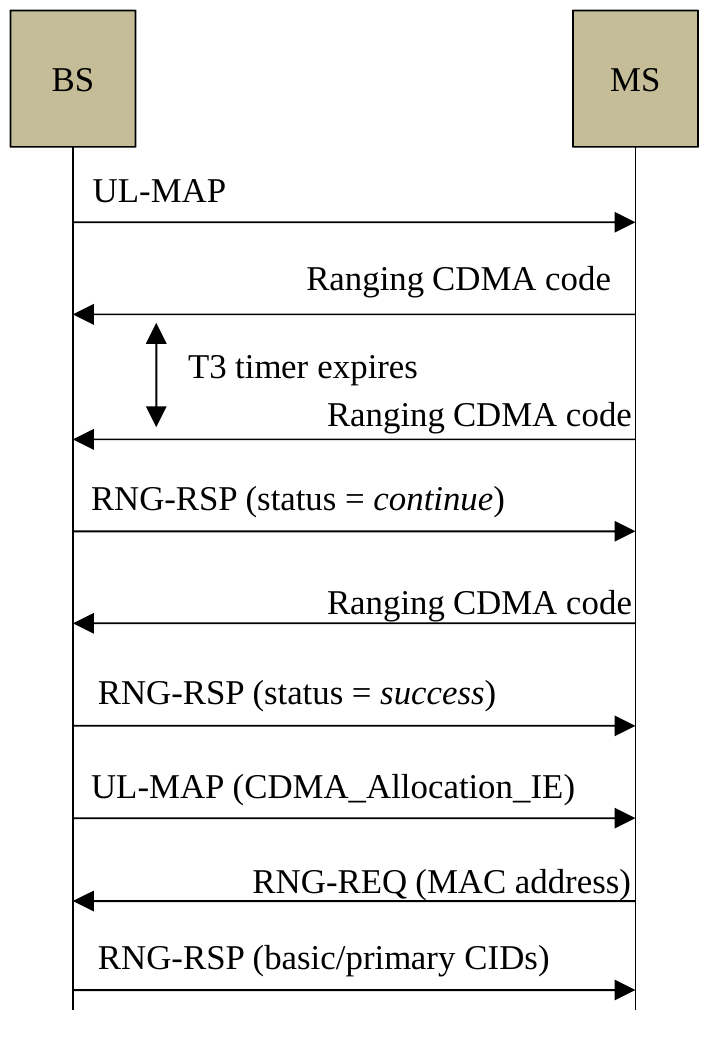}
\centering
\caption{Initial ranging procedure}
\end{figure}

Fig. 2 depicts the schematic of initial ranging procedure. For performing initial ranging, an MS first synchronizes to the DL and then acquires the UL channel characteristics through the UCD MAC management messages. The MS scans the UL-MAP message to find an initial ranging interval (which may consist of one or more transmission opportunities) that the BS has allocated in the UL sub-frame. When the initial ranging transmission opportunity occurs, the MS randomly selects a ranging CDMA code from a given initial ranging code-set, transmits the selected code at a certain power level (calculated with reference to its RSSI and the parameters specified in the DCD message), and waits for the ranging response (RNG-RSP) message that contains all the needed corrections (time, power, and frequency).

If the MS does not receive RNG-RSP from the BS within a certain time window (T3 timer), then the MS considers the previous ranging attempt to be unsuccessful and enters the collision-resolution stage. The MS sends a new ranging CDMA code at the next ranging opportunity, after an appropriate back-off delay determined using truncated binary exponential backoff algorithm, and with a one step higher power level. This process repeats until the maximum number of retries has been reached. On the other hand, if the MS receives a RNG-RSP, it will take subsequent actions based on status notifications indicated in the RNG-RSP. The three possible status notifications are as follows:
\begin{itemize}
  \item \emph{Continue}: If the MS receive RNG-RSP message containing the parameters of the code it has transmitted (OFDMA symbol number, subchannel, and the code itself) and status \emph{continue}, then the MS may considers the transmission attempt to be successful, but still requires further parameter adjustments. The MS thus adjusts its uplink transmission parameters as indicated in the RNG-RSP message and performs ranging by sending a ranging CDMA code randomly chosen from the initial ranging code-set on the suitably determined initial ranging opportunity.
  \item \emph{Success}: If the MS receive RNG-RSP message containing the parameters of the code it has transmitted and status \emph{success}, then the MS considers the transmission attempt to be successful. Upon receiving the \emph{success} status notification, the MS waits for uplink bandwidth allocation (CDMAAllocation\_IE in UL-MAP) to send ranging request (RNG-RREQ) message.  At this stage, the MS sends RNG-REQ message with its own MAC address and waits for the RNG-RSP which includes its MAC address as well as basic and primary Connection Identifier (CIDs).
  \item \emph{Abort}: If the MS receive RNG-RSP with status \emph{abort}, it abandons the ongoing ranging process.
\end{itemize}
Initial ranging procedure of the MS is considered to be successfully completed after it receives RNG-RSP which includes the valid \emph{basic} and \emph{primary} CIDs.

\section{Priority Initial Ranging}
In this Section, we first state the proposed mechanism and discuss the modifications that are deemed necessary in IEEE 802.16 standard to incorporate the proposed mechanism.

\subsection{The Proposed Mechanism}

The proposed priority network entry mechanism is discussed considering a scenario where two categories of MSs co-exist over a network. Let us say the MSs with ETS be high priority MSs and the rest be low priority MSs. The mechanism, however, can be extended easily for the cases where more than two priority categories exist.The proposed mechanism incorporates two approaches for access prioritization of high priority MSs. The first approach differentiates the size of starting RSSW of the high priority and low priority MSs. The differentiation is done in such a way that the RSSWs of high and low priority MSs remain inverse-proportional to their corresponding priorities as

\begin{equation}
\frac{I^h_{RSSW}}{I^l_{RSSW}}=\frac{x}{y},\ \ x < y
\end{equation}
\noindent where $I^h_{RSSW}$ and $I^l_{RSSW}$ are RSSWs for high priority and low priority MSs; and $x$ and $y$ are positive integers.

The second approach explicitly reserves a certain fraction (let us say $\alpha$) of available number of ranging codes to high priority MSs and allows low priority MSs use the remaining CDMA codes. It is noteworthy to mention that such an explicit CDMA ranging code reservation eliminates any collision chances among the inter-priority group MSs.

Note that, in 802.16 OFDMA network, each BS is assigned a subset of CDMA codes from the set of $256$ codes. That subset ranges from $S$ to $((S+N+M+L+O)mod\ 256)$ where $N$  codes are used for initial ranging, $M$  for periodic ranging,  $L$ for bandwidth request, and $O$  for handover ranging. Among $N$  initial ranging codes the proposed scheme reserves the first $\lceil N \cdot \alpha \rceil$  codes to be exclusively used by high priority MSs.

The proposed priority mechanism is exactly the same as the conventional initial ranging mechanism with an exception that the high priority and low priority MSs contend using the aforesaid priority differentiated contention parameters in terms of RSSW and CDMA codes.

\subsection{Modifications in UCD Message Format}

Recall that in order to perform initial ranging an MS first should synchronize to DL and then  acquires information about (i) initial ranging contention parameters  (number of initial ranging CDMA codes and minimum (or start)/maximum (or end) RSSW size for performing backoff) through the BS-broadcasted UCD management message, and (ii) available initial ranging resources (in terms of number of OFDMA symbols, number of OFDMA subchannels, OFDMA symbol offset, and OFDMA subchannel offset) through UL-MAP-IE (Uplink Interval Usage Code (UIUC) $= 12$).

In the proposed mechanism, high priority MSs and low priority MSs contend for initial ranging using different RSSW sizes and explicitly separated initial CDMA ranging codes. Hence, to incorporate the proposed mechanism in the IEEE 802.16 that RSSW differentiation and CDMA codes separation information should be included in the UCD message. Fig. 3 shows a modified UCD frame format with the three additional fields (marked in gray) along with the retained fields from the conventional UCD format [4, Section 6.3.2.3.3].

\begin{figure}[h]
\centering
\includegraphics[width=115mm, height=60mm]{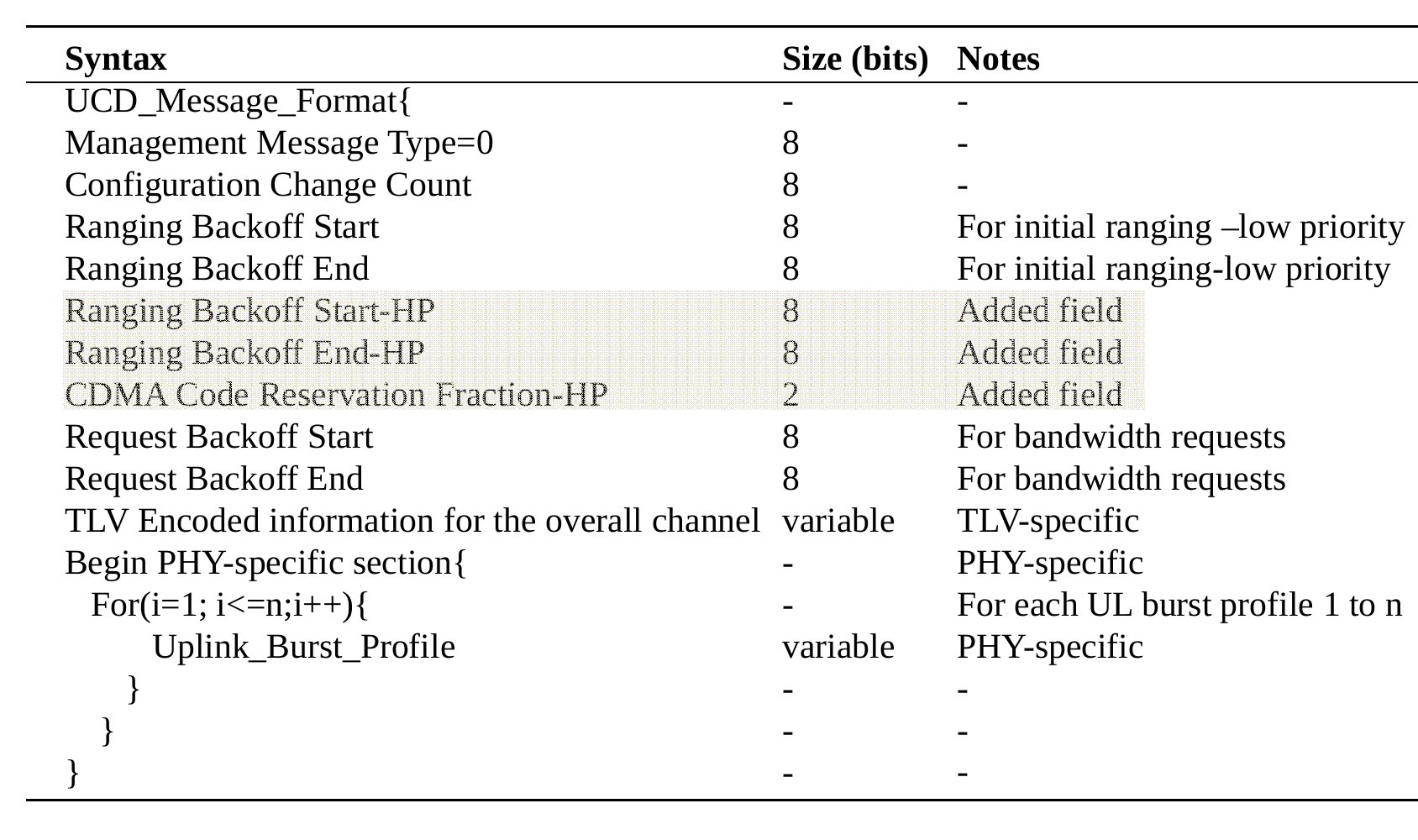}
\centering
\caption{A modified UCD message format}
\end{figure}

We consider the conventional \emph{Ranging Backoff Start} and \emph{Ranging Backoff End} fields are the RSSWs for low priority MSs, while the newly inserted \emph{Ranging Backoff Start\_HP} and \emph{Ranging Backoff End\_HP} are RSSWs for high priority MSs.  Both the newly inserted backoff start and end fields are $8$ bits long as the existing backoff start and end fields. On the other hand, the new field \emph{CDMA Code Reservation Fraction-HP} is $2$ bit long. It should be noted that the number of bits used to indicate \emph{CDMA Code Reservation Fraction-HP} determines the least count of the reserved CDMA codes for high priority MS. For example, if we consider the \emph{CDMA Code Reservation Fraction-HP} field to be of $2$ bits, then the least count is $8 (= 32/2^2)$ when the total number of initial ranging CDMA codes is $32$.

\section{Simulation Results and Discussion}

We conducted a series of simulations using our custom simulator developed in Matlab to examine the performance of the proposed initial ranging mechanism. In our simulation trials, we modeled the proposed contention based initial ranging process over a simplified network consisting of a single BS and increasing number of MSs (up to a certain number, let’s say $U$) that randomly arrive at the network (with a given arrival probability $p_a$) at discrete interval of simulation time (in unit of frame duration). Among the total arrived MSs, a certain fraction of MSs was high priority MSs, while the rest fraction was low priority MSs.

In our simulation, initially there were no MSs in the network. As the simulation time $i \in (1,2,\cdot \cdot \cdot,  n)$  was progressed, the number of MSs were increased in such a way that the number of arrived MSs $U_i$ until simulation time $i$  was equal to $U\cdot(1-(1-p_a)^i)$ for $i \ge 1$. Fig. 4(a) shows an example of a random arrivals of high priority and low priority MSs when $U$  is 200 and $p_a$  is $0.1$, while Fig 4(b) shows the average cumulative number of arrivals for $10$ different cases of $p_a$, $0.1$ to $1$ with the step size of $0.1$.

\begin{figure}[h]
\centering
\subfigure[]{%
\includegraphics[width=70mm, height=60mm]{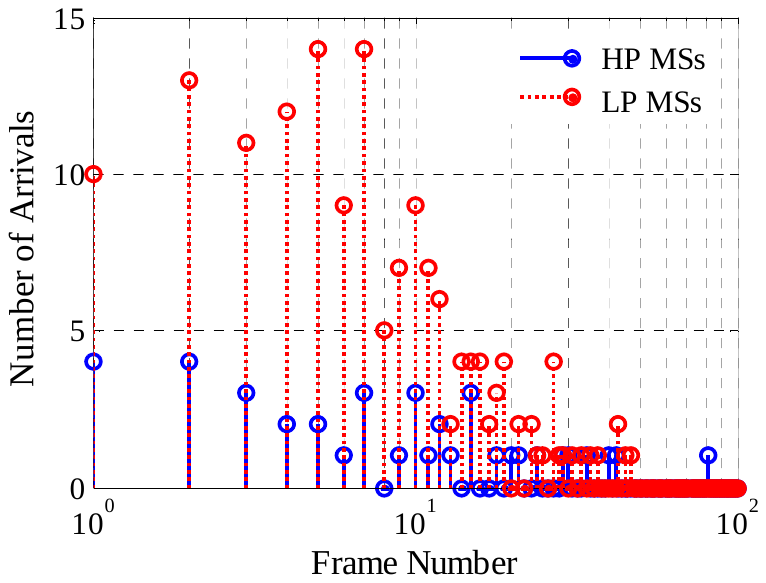}
\label{fig:subfigure1}
}
\subfigure[]{%
\includegraphics[width=70mm, height=60mm]{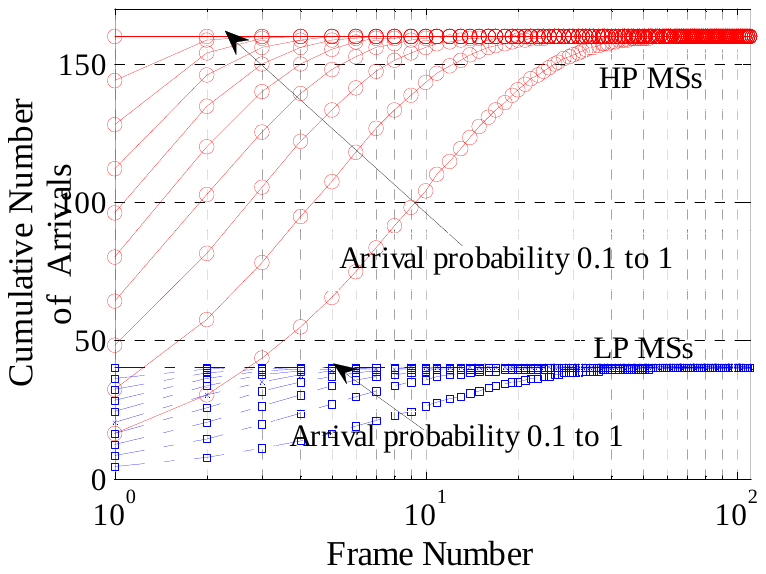}
\label{fig:subfigure2}
}
\caption{Arrival of MSs over the discrete simulation time in the unit of frame duration: (a) An example of a random arrival of MSs, and (b) Average cumulative number of arrived MSs}
\end{figure}

Table 1 shows the simulation parameters considered in this study including MS arrival parameters, total available ranging resources, and ranging contention parameters. It is worth mentioning that CDMA code detection error (false alarm and missed-detection) at BS increases with the increase in number of MSs transmitting at the same ranging opportunity \cite{Minn}. To reflect such physical layer effect in our MAC layer analysis, we have restricted the maximum number of MSs acceptable in a ranging opportunity (denoted as $\beta$) to four, though theoretically it could be up to $min(U,numder\ of\ ranging\  CDMA\ codes)$.

\begin{table}
\caption{Simulation Parameters}
\label{Tab1}
\begin{tabular}{ll}
\hline\noalign{\smallskip}
Parameters  &  Values   \\
\noalign{\smallskip}\hline\noalign{\smallskip}
Total number of MSs & 200\\
Arrival probability of MSs in a frame & variable: 0.1  to 1\\
Fraction of high priority MSs \\(i.e number of high priority MSs/total MSs) & 0.2 \\
Ranging opportunities per ranging channel    & 5 \\
Number of CDMA ranging codes & 32\\
Frame duration & 5 ms \\
T3 time & 20ms\\
Maximum number of MSs acceptable in\\ a ranging opportunity $\beta$ & 4 \\
CDMA code allocation (High Priority MSs) & variable: 25 \%, 50 \%, 75 \% \\
Initial RSSW & variable: 1 to 512 \\
Final RSSW & 1024 \\
\noalign{\smallskip}\hline
\end{tabular}
\end{table}

We first analyzed the performance of the proposed scheme in terms of success ratio of contending MSs considering two different cases: (a) Varying RSSWs of HP MSs and LP MSs for a given number of CDMA codes reserved to HP MSs, and (b) Varying number of CDMA codes reserved to HP MSs for given RSSWs of HP MSs and LP MSs. Note that success ratio is a ratio of number of MSs that are successful in completing their initial ranging to the total contending population. This ratio is always bounded within [$0, 1$].

\begin{figure}[h]
\centering
\subfigure[]{%
\includegraphics[width=70mm, height=60mm]{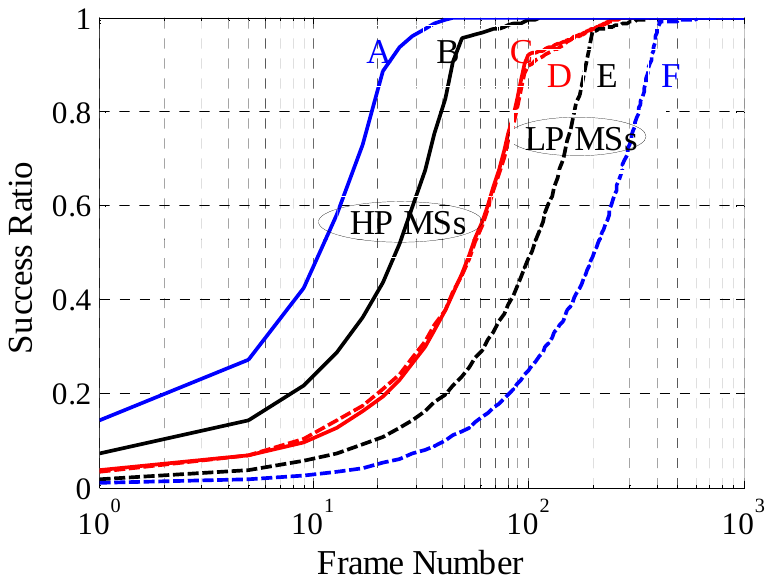}
\label{fig:subfigure5a}
}
\subfigure[]{%
\includegraphics[width=70mm, height=60mm]{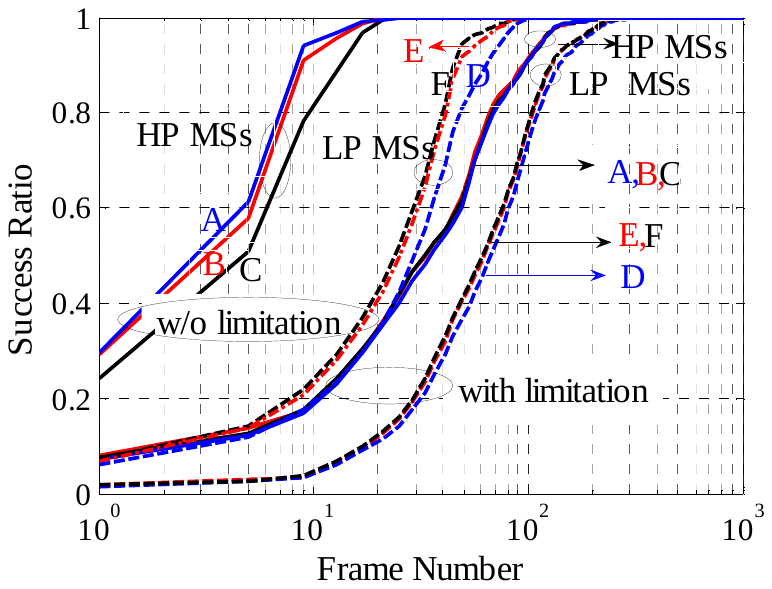}
\label{fig:subfigure5b}
}
\caption{Effect of differentiated contention parameters for high priority (HP) and low priority (LP) MSs in terms of their success ratio: (a) Effect of differentiated RSSWs, and (b) Effect of number of reserved CDMA codes. Legends in (a): (i) Curve pair CD: RSSW of HP MSs = $128$ and LP MSs = $128$, (ii) Curve pair BE: RSSWs of HP MSs = $64$ and LP MSs = $256$, and (iii) Curve pair AF: RSSWs of HP MSs = $32$ and LP MSs = $512$. Legends in (b): (i) Curve pair AD: CDMA code to HP MSs = $75\%$ and LP MSs = $25\%$, (ii) Curve pair BE: CDMA code to HP MSs = $50\%$ and LP MSs = $50\%$, and (iii) Curve pair CF: CDMA code to HP MSs = $25\%$ and LP MSs = $75\%$}
\end{figure}

Fig. 5(a) shows the success ratio of HP MSs and LP MSs over discrete simulation time (in unit of frame duration) for three different combinations of RSSWs of HP MSs and LP MSs, i.e., $(128, 128)$, $(64, 256)$, and $(32, 512)$, when $U = 200$,$p_a=1$, and the available $32$ CDMA codes were equally divided among the HP MSs and LP MSs. From the figure it is evident that the proposed mechanism maintains higher success ratio for HP MSs throughout the considered simulation time when the ratio of RSSWs of HP MSs to RSSWs of LP MSs is less than $1$. The smaller the RSSW ratio is, the higher is the success ratio of HP MSs. This can be jointly attributed to (i) higher access probability of HP MSs per ranging opportunity due to their relatively smaller RSSWs, and (ii) avoidance of collision likelihood among HP MSs and LP MSs due to their exclusively disjoint CDMA code set. Fig. 5(b) shows that for a given differentiated RSSWs of HP MSs and LP MSs, i.e. $(16, 64)$, no significant increase (decrease) in success ratio of HP MSs is possible by increasing (reducing) the number of CDMA code assigned to HP MSs from $50\%$ to $75\%$ $(25\%)$, especially when the physical layer detection failure is considered, i.e. for the case when $\beta$ is limited to $4$. However, for the ideal case, i.e. without any limitation in $\beta$, marginal increment (decrement) in success ratio of HP MSs is noticed when the number of CDMA code assigned to HP MSs is increased (decreased). With reference to Fig. 5, one can say that it is desirable to sufficiently differentiate RSSWs of HP MSs and LP MSs, while reserving small fraction of CDMA codes (for example $25\%$) to HP MSs.

Finding optimal RSSWs for HP MSs and LP MSs which can sufficiently meet Target Success Ratio Requirement (TSRR) of HP MSs (for example more than $99\%$) with the least possible sacrifice in LP MSs success ratio is vital for the best usage of our proposed ranging mechanism.

\begin{figure}[h]
\centering
\subfigure[]{%
\includegraphics[width=60mm, height=50mm]{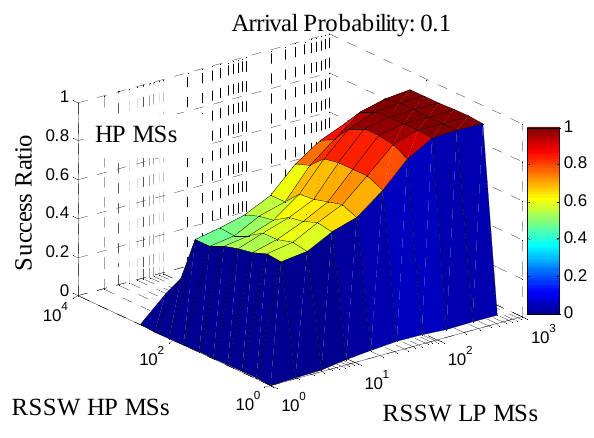}
\label{fig:subfigure5a}
}
\subfigure[]{%
\includegraphics[width=60mm, height=50mm]{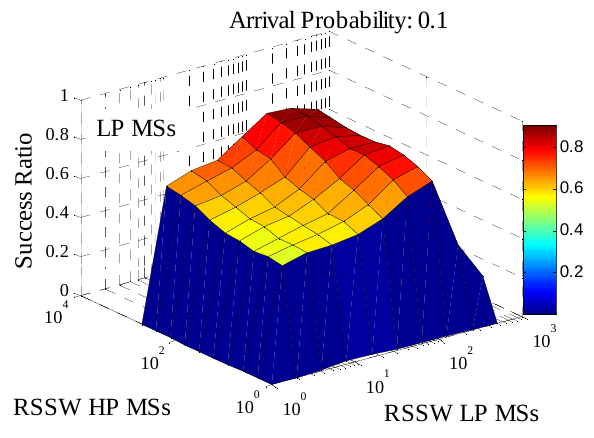}
\label{fig:subfigure5a}
}
\subfigure[]{%
\includegraphics[width=60mm, height=50mm]{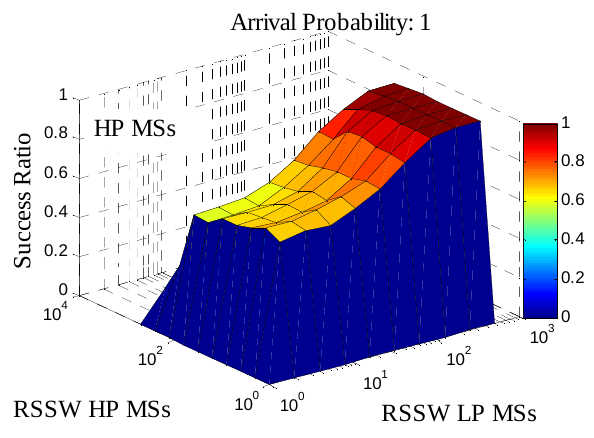}
\label{fig:subfigure5a}
}
\subfigure[]{%
\includegraphics[width=60mm, height=50mm]{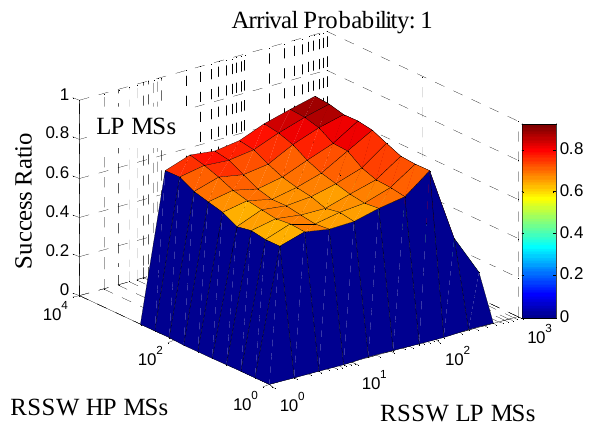}
\label{fig:subfigure5b}
}
\caption{Success ratio of MSs for varying RSSWs when only a quarter of the available CDMA codes were reserved for HP MSs: (a) HP Mss (arrival probability = $0.1$), (b) LP MSs (arrival probability = $0.1$), (c) HP MSs (arrival probability = $1$), and (d) LP MSs (arrival probability = $1$)}
\end{figure}

Fig. 6 shows success ratio of HP MSs and LP MSs within 0.5 s (equivalent to 100 frames) considering the two different cases of arrival probabilities, i.e  = 0.1 and  = 1. We found that (16, 128) and (32, 128) were optimal RSSWs for HP MSs and LP MSs for  = 0.1 and  = 1, respectively, when the TSRR $\ge 98\%$. With the optimal RSSWs settings, success ratio of HP MSs and LP MSs were $99.9\%$ and $83.72\%$ in the first case, while $98.95\%$ and $88.69\%$ in the second case. With reference to Fig. 6 it can be said that desired TSRR for high priority MSs can be guaranteed by selecting optimal RSSWs and CDMA-code reservation fraction when the target network synopsis (for example expected number of MSs in the network, fractions of HP MSs, MSs arrival rate) is known.

\section{Conclusions}
Prioritized network entry could be one of the most important procedures to provide preferential network connectivity to MSs relative to their associated priorities. However, until recently, it has not been widely conceived, studied, and analyzed. In this paper, we have presented a simple mechanism to support priority differentiated network entry in IEEE 802.16 OFDMA networks. Through rigorous simulations, we have analyzed performance of the proposed mechanism and showed that for a given network conditions (for example number of MSs, fractions of HP MSs, MSs arrival rate) desired target success ratio for high priority MSs can be guaranteed by properly choosing the size of RSSWs and the CDMA-code reservation factor.



\begin{thebibliography}{}

\bibitem{Ahmadi}
Ahmadi S. Mobile WiMAX: a system approach to understanding IEEE 802.16m radio access technology. Academic Press, 2010.

\bibitem{Taha}
Taha AM, Ali NA, Hassanein HS. LTE,LTE-advanced and WiMAX: towards IMT-advanced networks. John Wiley and Sons, 2012.

\bibitem{Andrews}
Andrews JG, Gosh A, Muhamed R. Fundamentals of WiMAX: understanding broadband wireless networking. Prentice Hall, 2007.

\bibitem{standard}
IEEE Std. 802.16-2012, IEEE standard for local and metropolitian area networks- part 16: air interface for broadband wireless access systems (August 2012)

\bibitem{Cicconetti}
Cicconetti C, Lenzini L, Mingozzi E, Eklund C. Quality of service support in IEEE 802.16 networks. IEEE Network 2006; 20: 50-55.

\bibitem{Zhang}
Zhang H, Wang X, Qin ZB, Kuo GS, Bohnert TM. Adaptive QoS provision for IEEE 802.16e BWA networks based on cross-layer design. Eurasip Journal on Wireless Communications and Networking 2011, 2011:69

\bibitem{Folts}
Folts H. Standard initiatives for emergency telecommunications service. IEEE Communication Magazine 2002; 40: 102-107.

\bibitem{Boone}
Boone P, Barbeau M, Kranakis E. Prioritized access for emergency stations in next generation broadband wireless networks. Proceedings of the annual conference on communication networks and services research, 2010. p. 319-326.

\bibitem{Zhu}
Zhu H, Ming L, Chalmtac I, Prabhakaran B. A survery of quality of service in IEEE 802.11 networks. IEEE Wireless communication 2004; 11: 6-14.

\bibitem{weblink}
Initial ranging for priority access in IEEE 802.16.1a, IEEE 802.16 broadband wireless access working group (2012), https://mentor.ieee.org/802.16/documents

\bibitem{Minn}
Minn H, Fu X. A new ranging method for OFDMA systems. Proceedings of IEEE Globecom, 2005. p. 1435-1440.

\end{thebibliography}
\end{document}